\documentclass[useAMS,usenatbib]{mn2e}

\usepackage{graphicx}  

\title[Raman Scattered Ne VII 973 in V1016~Cyg]{Raman Scattered 
Ne VII$\lambda$973 at 4881 \AA\
in the Symbiotic Star V1016~Cygni}
\author[Hee-Won Lee, Jeong-Eun Heo, and Byeong-Cheol Lee]{Hee-Won Lee,$^{1}$\thanks{E-mail:
hwlee@sejong.ac.kr (HWL)} Jeong-Eun Heo,$^1$\thanks{E-mail: jeung6145@gmail.com} and 
Byeong-Cheol Lee$^{2}$\thanks{E-mail: bclee@kasi.re.kr}\\
$^{1}$Department of Astronomy and Space Science, Sejong University, Seoul, 143-747, Korea\\
$^{2}$Korea Astronomy and Space Science Institute, Daejeon, Korea}

\begin{document}

\date{Accepted 1988 December 15. Received 1988 December 14; in original form 1988 October 11}

\pagerange{\pageref{firstpage}--\pageref{lastpage}} \pubyear{2002}

\maketitle

\label{firstpage}

\begin{abstract}
We present the high resolution spectra of the symbiotic star V1016~Cygni obtained with
 the Bohyunsan Optical Echelle Spectrograph in 2003 and 2005, from which we 
 find a broad emission feature at 4881 \AA. We propose that  this broad feature
is formed from Raman scattering of Ne~VII$\lambda973$ by atomic hydrogen.
Thus far, the detection of Raman scattered lines by atomic hydrogen is limited
to O~VI$\lambda\lambda$1032, 1038 and He~II$\lambda\lambda$940, 972 and 1025.
With the adoption of the center wavelength 973.302 \AA\ of Ne~VII$\lambda$973  
and consideration of the air refractive index of $n_{air}=1.000279348$, 
the atomic line center of the Raman scattered Ne~VII feature is determined to 
be 4880.53~\AA. 
The total cross section at the line center of Ne~VII$\lambda973$ is computed to be $2.62\times
10^{-22}{\rm\ cm^2}$ with the branching ratio of 0.17.
We perform Monte Carlo simulations to fit the Raman scattered Ne~VII$\lambda973$.
Assuming that the Ne~VII and He~II emission
regions share the same kinematics with respect to the neutral
scattering region, we find that the Raman scattered He~II$\lambda$972 at 4850 \AA\ 
and Ne~VII$\lambda$973 at 4881 \AA\ are 
excellently fitted. We also propose that 
the He~II and Ne~VII
emission regions were stationary with respect to the H~I region in 2003 but that
they were receding from it with a velocity $\sim 20{\rm\ km\ s^{-1}}$ in 2005.

\end{abstract}


\section{Introduction}

Symbiotic stars are believed to be wide binary systems consisting of
a hot component, usually a white dwarf, and a mass losing giant 
(e.g. \citealt{ke86}, \citealt{bel00}). Because of the size of the giant component,
their orbital periods range from a few hundred days to several decades
(e.g. \citealt{mi12}).
Symbiotic stars exhibit prominent emission lines indicative of a wide range of ionization including [SII] and
O~VI. Some fraction of the slow stellar wind from the giant component
is accreted to the hot component, which leads to various activities 
including occasional eruptions and strong emission lines (e.g.
\citealt{ib96}, \citealt{lp99}).
\cite{ma98} 
performed {\it Smoothed Particle Hydrodynamical} 
computations for binary systems of a mass losing giant and a white dwarf
to show that an accretion disk may be formed through accretion of the stellar
wind. However, it is still an unresolved issue
whether an accretion disk is formed in general in symbiotic stars (e.g. \citealt{sok01}, \citealt{nu03}).

Symbiotic stars are classified 
into 'S' type and 'D' type, where the latter exhibit an IR excess indicating the existence
of a warm dust component surrounding the binary system.   
It appears that D-type symbiotic
stars possess multiple dust shells with temperatures of $\simeq 1000{\rm\ K}$ and
$\simeq 400{\rm\ K}$ (\citealt{an10}). Many symbiotic stars are
also X-ray emitters exhibiting a large range of hardness. \cite{lu13} classified
X-ray emitting symbiotic stars into types $\alpha, \beta,\gamma$ and  $\delta$ 
depending on the hardness with the $\alpha$ type being supersoft X-ray sources
and the $\delta$ type being highly absorbed hard X-ray emitters. They discussed many astrophysical
processes including thermonuclear shell burning, wind collision and boundary layer
in the accretion disk.

In symbiotic stars unique and useful spectroscopic diagnostics are provided from 
Raman scattering by atomic
hydrogen. In this scattering process, a far UV photon blueward of 
Ly$\alpha$ is incident upon a hydrogen atom in the ground state. Subsequently the hydrogen
atom de-excites into $2s$ state re-emitting an optical Raman-scattered photon.
The first identification of Raman scattering in symbiotic stars
was made by \cite{sc89}, who proposed that the broad emission features at
6825 \AA\ and 7088 \AA\ usually observed in many symbiotic stars are Raman
scattered O~VI$\lambda$ 1032 and O~VI$\lambda$ 1038.
 The operation of Raman
scattering requires the coexistence of a highly thick H~I region and a strong far UV emission
region, which is ideally met in symbiotic stars. Recently Raman scattered O~VI 1032 and 1038
features were found in the B[e] star LHA 115-S 18 (\citealt{to12}). 

Raman spectroscopy is very important in probing the mass loss and mass transfer
processes in symbiotic stars (e.g. \citealt{sc89}, \citealt{lp99}).
Adopting an emission region that is in a Keplerian motion around the hot component, 
\citet{lk07} successfully fitted the double peaked profiles of the Raman scattered
O~VI$\lambda$1032 of the symbiotic stars HM Sge and V1016 Cygni. The representative
Keplerian velocity scale $\sim 30{\rm\ km\ s^{-1}}$ of the emission regions 
in these two symbiotic stars implies that
the major O~VI emission region is located roughly within one astronomical unit
from the hot component. It should be also noted that a double peak profile in Raman
scattered O~VI can be produced from purely geometric effects, where the scattering
region coincides with the slowly expanding stellar wind around the giant (e.g.
\citealt{hh97}, \citealt{sch92}).
The mass loss rate of the cool component of V1016~Cyg has been deduced from a
photoionization model calculation combined with
the measurement of the center shift of Raman scattered He~II by \cite{jl04}. 
 
Additional Raman scattered features were found and proposed by \cite{vg93}, 
who discussed
the plausibility of Raman scattering for various far UV emission lines 
including He~II and
C~III. In particular, he reported 
the detection of the Raman scattered He~II features blueward
of H$\beta$ and H$\gamma$ in the symbiotic star RR Telescopii. 
These features
are also found in the symbiotic stars HM~Sge and V1016~Cyg (e.g. \citealt{bi04}, 
\citealt{jl04}, \citealt{lee12}).

In a Be-like isoelectronic sequence, Ne~VII is a very highly 
ionized species with the ionization potential of 207.271 eV (e.g.
\citealt{kr13}). Because of this high ionization potential, Ne~VII lines are
reported only in extremely hot stellar objects and active galactic nuclei.
\cite{we04} found Ne~VII$\lambda$973 absorption lines in the spectra of
PG~1159 stars obtained with the {\it
Far Ultraviolet Spectroscopic Explorer (FUSE)}.
\cite{yo05} proposed the existence of Ne~VII$\lambda$973 in the FUSE spectrum 
of the symbiotic star AG Draconis. In the Seyfert 1 galaxy NGC~5548 \cite{ka95}
identified Ne~VII/Ne~VIII blend at 88~\AA\ in the extreme UV spectrum.

V1016~Cyg is a 'D' type symbiotic star with the cool component 
being a Mira type variable. It is also a symbiotic nova that underwent 
a nova-like outburst in 1964 (\citealt{mc65}). According to the investigation
of \cite{mu91}, 'D' type symbiotic novae including HM~Sge, RR~Tel and V1016~Cyg 
form the subgroup of symbiotic stars with the hot components
characterized by the highest temperature and luminosity. 
V1016~Cyg being of higher temperature than AG Dra, this leads to an interesting 
possibility that V1016~Cyg may show the emission line Ne~VII$\lambda$973.

In this paper, we present the high resolution spectra of V1016~Cyg, in which
a broad emission feature at 4881 \AA\ is found redward of H$\beta$. We propose this 
broad feature is formed through Raman scattering of the far UV emission
line Ne~VII$\lambda$973 in the thick H~I region.

\section{Observation}
\subsection{High resolution spectroscopy of V1016~Cyg}
\begin{figure}
\vspace{20pt}
\includegraphics[scale=0.67]{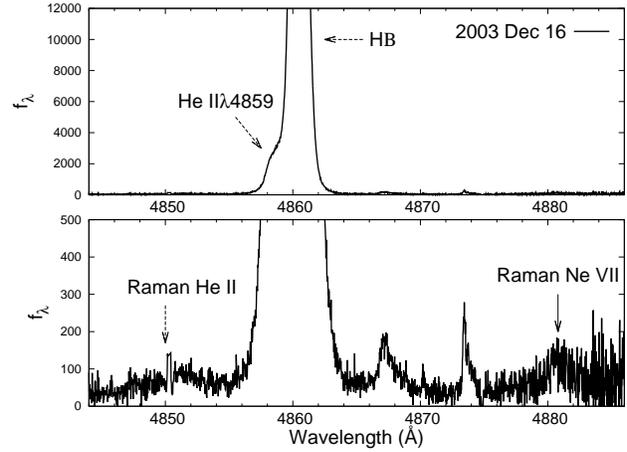}
 \caption{The high resolution spectrum of V1016~Cyg near H$\beta$ obtained 
 with BOES on 2003 December 16. The lower panel is a blow-up version of the
 upper panel by a factor 24. The vertical axis shows the CCD counts.
 The broad feature around 4882 \AA\ marked by
 the solid vertical arrow shown in the bottom panel is proposed to be 
 Raman scattered Ne~VII$\lambda$973 by
 atomic hydrogen.}
 \label{boes_fig1}
\end{figure}

\begin{figure}
\vspace{20pt}
\includegraphics[scale=0.67]{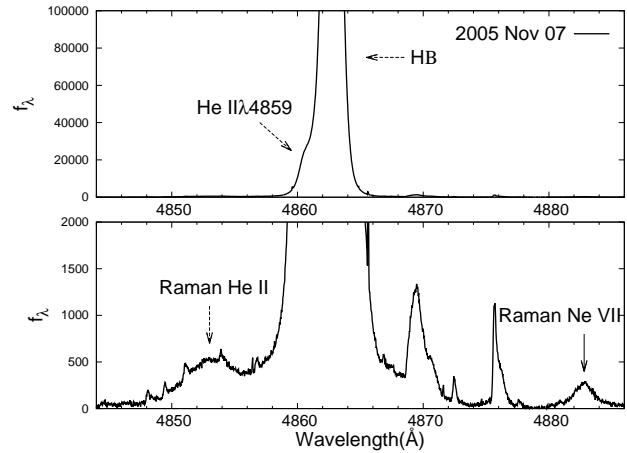}
 \caption{The high resolution spectrum of V1016~Cyg obtained 
 with BOES on 2005 November 7. The lower panel is a blow-up version of the upper
 panel by a factor 50.  The vertical axis shows the CCD counts.
The same broad feature is found around 4883 \AA\,
 which is marked by the solid vertical arrow shown in the bottom panel.}
 \label{boes_fig2}
\end{figure}

Fig.~\ref{boes_fig1} shows a high resolution 
spectrum of V1016~Cyg obtained with
the Bohyunsan Optical Echelle Spectrograph (BOES) installed on the 1.8 m
optical telescope at Mt. Bohyun on the night of 2003 December 16.
In Fig.~\ref{boes_fig2}, we show another BOES spectrum of V1016~Cyg obtained on 2005 November 7. 
For the spectrum of 2003, the 200 $\mu$m optical fiber was used to yield the spectroscopic
resolution of $\simeq 44,000$. On the other hand,
the 2005 spectrum was obtained using the 300 $\mu$m optical fiber achieving spectroscopic
resolution of $\simeq 30,000$. The exposure time was 3000 seconds and 7200 seconds
for the spectrum of 2003 and that of 2005, respectively. In Fig.~\ref{boes_fig1},
the lower panel is a blow-up version of the upper panel by a factor 24.
In a similar way, in Fig.~\ref{boes_fig2} the scale of the lower panel
is 50 times smaller than that of the upper panel.

\begin{figure}
\vspace{20pt}
\includegraphics[scale=0.67]{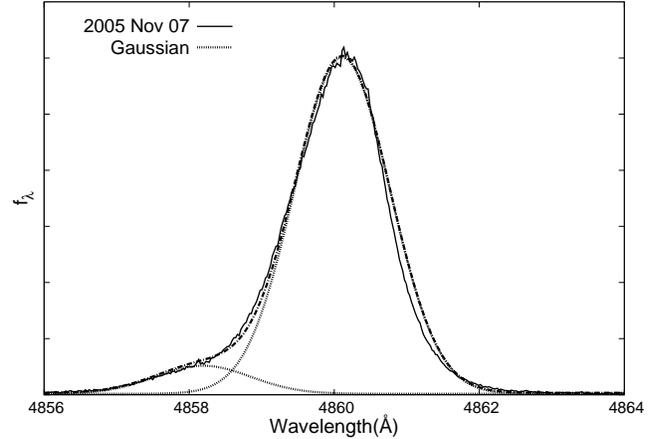}
 \caption{Profile fit of He~II$\lambda$4859 and H$\beta$ using 
a single Gaussian function for the spectrum obtained on the night of 
2005 November 7 with the exposure time of 300 seconds. Gaussian fitting 
functions are shown by dotted lines and the BOES
 spectrum is represented by a solid line. The BOES data are fitted by
combining the two Gaussian functions, which is represented by a
dot-dashed line.
 The observed line center of He~II$\lambda$4859 is 4858.57 \AA\ and
 that of H$\beta$ is 4860.54 \AA.}
 \label{gas_fit1}
\end{figure}

In the 2005 spectrum, the emission line
H$\beta$ is saturated in order to allow a close investigation of
the weak features around H$\beta$. We also obtained another spectrum with a short exposure time
of 300 seconds, which is shown in Fig.~\ref{gas_fit1}. 
As in the analysis of \cite{jl04}
we find Raman scattered He~II$\lambda$972 at 4850 \AA\ in the two long exposure spectra obtained in 2003 and 2005, 
which is very broad and formed in the blue wing part of H$\beta$.

Because of energy conservation, the wavelength $\lambda_{RV}$ of the Raman scattered
radiation in vacuum is related to that of the far UV incident radiation $\lambda_i$ by the
equation
\begin{equation}
\lambda_{RV}={\lambda_{Ly\alpha}\lambda_i\over \lambda_{Ly\alpha}-\lambda_i},
\label{raman_length}
\end{equation}
where $\lambda_{Ly\alpha}=1215.67{\rm\ \AA}$ is the line center wavelength 
of hydrogen Ly$\alpha$.  With the refractive index
of air $n_{air}=1.000279348$, the observed wavelength will be reduced to
\begin{equation}
\lambda_R=\lambda_{RV}/n_{air}.
\end{equation}
In particular, the incident wavelength $\lambda_i=
972.112{\rm\ \AA}$ for He~II$\lambda$972 gives rise to the Raman scattered
He~II at $\lambda_{RV, HeII}=4852.098{\rm\ \AA}$ (\citealt{jl04}), which yields the observed
wavelength of $\lambda_{R, HeII}=4850.743 {\rm\ \AA}$.

In Fig~\ref{boes_fig1}, the broad
Raman scattered He~II feature has the observed line center at 
$\lambda_{R, HeII}^{obs}=4850.63{\rm\ \AA}$, which is blueshifted from $\lambda_{R, HeII}$ by
$\Delta\lambda=0.11{\rm\ \AA}$ from the expected center wavelength. On the other hand, 
in Fig.~\ref{boes_fig2}, we find that $\lambda_{R, HeII}^{obs}=4852.95{\rm\ \AA}$.
  This deviation
can be explained that the He~II emission region is receding from
the neutral H~I region. More detailed investigation on the kinematics
of the emission region is described in the subsequent section. 
However, 
line center shift can also occur due to the varying Raman conversion rate
dependent on the wavelength (\citealt{jl04}), which necessitates Monte Carlo
simulations for more accurate determination of the kinematics of the He~II
emission region relative to the H~I region.

Eq.~(\ref{raman_length}) also leads to the following relation
\begin{equation}
{\Delta\lambda_{R}\over\lambda_{R}}
=\left({\lambda_{R}\over\lambda_i}\right)
{\Delta\lambda_i\over\lambda_i},
\label{profile_br}
\end{equation}
which results in Raman scattered features with a significantly broadened profile.
In particular, the Raman scattered features around H$\beta$ are broadened  
by a factor $\lambda_{R}/\lambda_i\simeq 5$, because $\lambda_i\simeq 972{\rm\ \AA}$
and $\lambda_R\simeq 4861{\rm\ \AA}$.

Using a Gaussian function given by
\begin{equation}
F(\lambda)=F_0\exp[(\lambda-\lambda_0)^2/\Delta\lambda^2],
\label{gaussian}
\end{equation}
we perform a line profile fitting analysis 
to determine the observed centers of the emission lines He~II$\lambda$4859 and H$\beta$   
in the 2005 spectrum of V1016~Cyg. 
Because of the saturation of H$\beta$, we use 
another spectrum obtained on the same night with an exposure time of 300 seconds.
In Fig.~\ref{gas_fit1}, we show the result, which is also summarized in Table~1.

\begin{figure}
\vspace{20pt}
\includegraphics[scale=0.67]{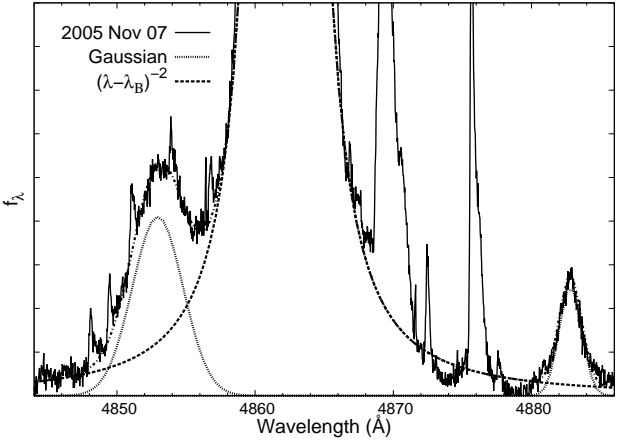}
 \caption{Profile fit of Raman scattered He~II$\lambda$972 and a broad feature at 4883 \AA\ 
using a single Gaussian function
for the spectrum obtained on the night of 2005 November 7 with the exposure time 
of 7200 seconds. For the H$\beta$ wing the profile function proportional 
to $(\lambda-\lambda_\beta)^{-2}$ is used for fitting, which is shown by a thick dotted line.
The BOES spectrum is shown by a solid line and the two single Gaussian functions are represented
by a dotted line. The dot-dashed line shows the combined fit of the two Gaussians and H$\beta$ wing.
The observed line center of Raman scattered He~II$\lambda$972 is 4852.95 \AA\
 and that of Raman scattered Ne~VII$\lambda$973 is 4882.82 \AA.}
 \label{gas_fit2}
\end{figure}

In a similar way, the Raman scattered He~II$\lambda$972 and the broad feature at 4883 \AA\
are fitted for the spectrum with exposure time of 7200 seconds, of which the result is shown in Fig.~\ref{gas_fit2}. For this, the broad H$\beta$ wings
are fitted using a profile $(\lambda-\lambda_\beta)^{-2}$, where $\lambda_\beta$ is the wavelength of H$\beta$.
In the case of the 2003 spectrum, the profile
fitting was done by \cite{jl04} except the broad feature at 4881 \AA.
In this work, the observed line center of this feature in the 2003 spectrum
is determined to be 4880.80 \AA. 
The result of our single Gaussian fit to the 2003 spectrum is also summarized in Table 1.

\subsection{Raman scattered Ne~VII$\lambda$973}

Both in Figs.~\ref{boes_fig1} and \ref{boes_fig2}, the broad emission feature marked by
a solid vertical arrow is noticeable. This feature in the 2003 spectrum is very weak with the
observed center at 4880.80 \AA\
but it is significantly clear at 4882.82 \AA\ in the spectrum of 2005 
with much longer exposure time. We propose that
this broad feature is formed through Raman scattering of Ne~VII$\lambda$973 by atomic
hydrogen.
 
Ne~VII$\lambda973$ is formed through recombination of Ne~VIII accompanied by a
radiative transition from $2p^2\ ^1D_2$ to $2s2p\ ^1P_1$ (e.g. \citealt{yo05}). This line
was also found in the high resolution coronal spectrum of the sun obtained with the
Solar Ultraviolet Measurement of Emitted Radiation (SUMER) on board the {\it Solar
and Heliospheric Observatory, SOHO} (\citealt{fe97}). \cite{he05}
proposed that Ne~VII$\lambda$973 is responsible for the strong P~Cygni features 
observed at around 975 \AA\ for
a number of hot evolved stars including A78, NGC 2371 and K1-16.

The atomic data for Be-isoelectronic atoms can be found in the work of \cite{ed83},
who compared the theoretical values with the experimental data.
He recommended the values of 214954 cm$^{-1}$ and 317954 cm$^{-1}$ for the lower
and the higher energy levels respectively, from which 
the center wavelength of Ne~VII$\lambda$973 is obtained 
to be 973.350 \AA. However, according to \cite{kr13}, the energy values of the lower
and the higher levels are 214951.6 cm$^{-1}$ and 317694.6 cm$^{-1}$ respectively, 
resulting in the central wavelength of 973.302 \AA. This result is based on the
previous work of \cite{kr06}. In this work we adopt this value as the line center
wavelength $\lambda_0=973.302{\rm\ \AA}$ of Ne~VII$\lambda$973.

A direct substitution of the Ne~VII line center wavelength into Eq.~(\ref{raman_length}) 
yields the vacuum line center wavelength of the Raman scattered Ne~VII feature, which is
given by 
$\lambda_{Ne}^{RV}= 4881.89{\rm\ \AA}$.
Considering the refractive index of air, this vacuum wavelength
is reduced to
\begin{equation}
\lambda_{Ne}^{R}= 4880.53{\rm\ \AA},
\end{equation}
from which we propose to call this broad feature 'Raman scattered Ne~VII$\lambda$973'.
In Section 3, we perform Monte Carlo simulations to fit the profile, from which
we may deduce useful constraints on the strength and profile of the unobserved 
Ne~VII$\lambda$973.

This line center value $\lambda_{Ne}^{R}$ is shorter than the observed center wavelength 
$\lambda^{obs}_{R} =4880.80{\rm\ \AA}$ by a small amount of $\Delta\lambda=0.27{\rm\ \AA}$ 
in the 2003 spectrum. In the case of the 2005 spectrum, the corresponding values are  
$\lambda^{obs}_{R}
=4882.82{\rm\ \AA}$, and  $\Delta\lambda=2.29{\rm\ \AA}$. This feature is
quite broad, which is consistent with the primary characteristic of a Raman scattered feature. 
  A similar spectroscopic behavior is seen in the case of
Raman scattered He~II$\lambda$972. In the 2003 data, the observed line center of
Raman scattered He~II$\lambda$972 is almost coincident with that expected from the
atomic line center. However, in the 2005 spectrum it is redshifted by an amount
exceeding 2 \AA. Because both He~II and Ne~VII
are highly ionized species, it is quite plausible that their emission regions 
almost coincide sharing a similar kinematics with respect to the neutral region. 
Because the profiles of Raman scattered features are dominantly affected by a relative
motion between the far UV emission source and the H~I region and little influenced
by the observer's sightline, the difference in the spectra of 2003 and 2005
implies that there was a significant change in the far UV emission region and/or in the H~I region
of V1016~Cyg.

\begin{table*}
\centering
\caption{Observed wavelengths of emission lines and Raman scattered features in the two
high resolution spectra of V1016~Cyg}
\begin{tabular}{ccccc} \hline
\hline
Line & Raman He~II$\lambda$972 at 4850&  He~II$\lambda$4859 & H$\beta$ & Raman Ne~VII$\lambda$973 at 4881\\ 
 Observed Wavelength & $\lambda_{4850}^{obs}$ & $\lambda_{4859}^{obs}$ & $\lambda_\beta^{obs}$ & $\lambda_{4881}^{obs}$ \\
 \hline
2003 Dec 16 & 4850.63 & 4858.57 & 4860.54 & 4880.80 \\ 
2005 Nov 07 & 4852.95 & 4858.19 & 4860.11 & 4882.82 \\ \hline
Atomic Line Center & 4850.74 & 4859.32 & 4861.28  & 4880.53 \\ \hline
\end{tabular}
\end{table*}

\subsection{Temporal Variation of the Kinematics of the Far UV Emission Regions}

We investigate the temporal change in relative velocities 
of the He~II emission region and Ne~VII
emission region with respect to the neutral H~I region, 
which may have happened during the two year period. 
In this work, the atomic line center of H$\beta$ is set to be $\lambda_\beta=4861.28{\rm\ \AA}$ 
in air and we assume that the Balmer emission region suffered no significant changes 
in kinematics. In Table~2, 
we provide the observed wavelengths corrected for the rest frame of H$\beta$ emission region. 
It is unclear whether the Balmer emission region may represent the systemic
velocity of V1016~Cyg due to some variability in this wide binary system. However, a 
reference system is useful in discussion of the relative motion of the He~II and Ne~VII 
emission regions and we take the Balmer emission region as a kinematic reference.

In Table~2, we also show the line center wavelengths of Raman scattered He~II$\lambda$972,
He~II$\lambda$4859 and Raman scattered Ne~VII$\lambda$973 corrected for the
rest frame of H$\beta$. In the year 2003, the corrected line center of
He~II$\lambda$4859 almost coincides with that of the atomic line center, implying
that at that time the He~II emission region is almost stationary with respect to
the Balmer emission region. Despite this, Raman scattered He~II$\lambda$972 is redshifted
by an amount 0.63 \AA. This redward shift is attributed to the atomic physical effect, which
was pointed out by \cite{jl04}. According to them, the
scattering cross section and branching ratio are increasing functions of wavelength
locally around the line center of He~II$\lambda$972. This enhances Raman scattered radiation
redward of He~II$\lambda$972, resulting in an overall redward shift of the Raman scattered 
feature. The exact amount of wavelength shift
depends on the H~I column density $N_{HI}$. Their Monte Carlo simulations yielded the value of $N_{HI}=1.2\times 10^{21}{\rm\ cm^{-2}}$ for V1016~Cyg.

  However, the wavelength shift of 3.38 \AA\ of Raman scattered He~II$\lambda$972
shown in the 2005 spectrum is too large to be accounted for by
the effect of atomic physics. This implies that the He~II emission region was receding from
the H~I region at the time of the observation in 2005. Both the Doppler shift and the
atomic physics affect the wavelength shift in a complicated way so that one requires
Monte Carlo simulations in order to determine the kinematics of the He~II and Ne~VII
emission regions. Exactly the same argument applies to the Raman scattered Ne~VII$\lambda$973,
which constitutes the content of the following section. 


\begin{table*}
\centering
\caption{Difference of the observed line wavelength from that of the observed H$\beta$}
\begin{tabular}{cccc} \hline
\hline
Line & Raman He~II$\lambda972 at 4850$ &He~II$\lambda$4859 & Raman Ne~VII$\lambda$973 at 4881 \\ 
Wavelength Corrected & $\lambda_{4850}$ & $\lambda_{4859}$ & $\lambda_{4881}$ \\ \hline
2003 Dec 16 & 4851.37 & 4859.31  & 4881.54\\ 
2005 Nov 07 & 4854.12 & 4859.36  & 4883.99\\ \hline
\end{tabular}
\end{table*}


\section{Monte Carlo Simulations}

\subsection{Cross sections and branching ratios}

The scattering
cross section is computed from the second order time-dependent perturbation
theory described in many textbooks on quantum mechanics including \cite{bs57}. 
The cross section
is given by the Kramers-Heisenberg formula, where the matrix elements
are provided in a straightforward way in the case of a single electron
atom such as hydrogen. Blueward of Ly$\gamma$, there are 4 scattering channels
depending on the principal quantum number of the final state of the hydrogen atom,
where the final state can be the $s$ states of $n=1$ and $n=2$, and the
$s$ and $d$ states of $n=3$ and 4. Redward of Ly$\gamma$,
states with $n=4$ are excluded from the final state, leaving only 3 scattering
channels.
When the final state is $2s$, then we have optical scattered radiation near
H$\beta$. Final states with $n=3$ and $n=4$ lead to Raman scattered radiation in the IR region.
The total scattering cross section $\sigma_{tot}$ and the branching ratios
around Ly$\gamma$ were computed and presented by \cite{lj06} in their analysis
of the Raman scattered He~II$\lambda$972 in the young planetary nebula IC~5117.

\begin{figure}
\includegraphics[scale=0.62]{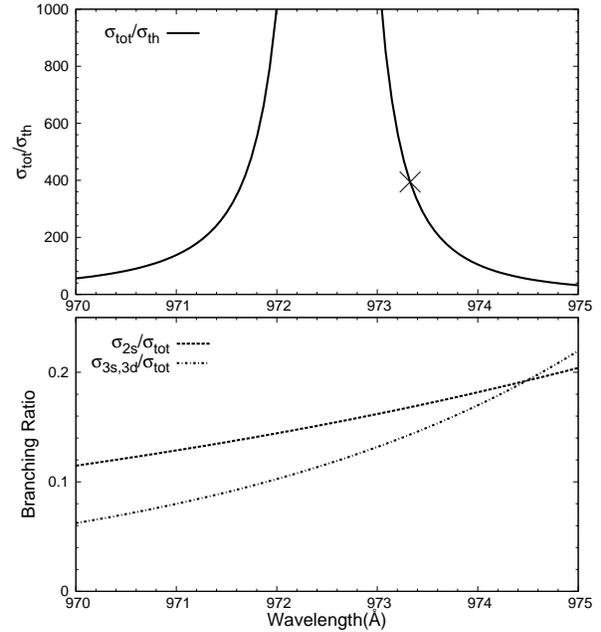}
 \caption{Atomic data for Rayleigh and Raman scattering around Ly$\gamma$. In the upper panel,
the total cross section $\sigma_{tot}$ of Rayleigh and Raman scattering near Ly$\gamma$ is shown by
a solid line. The cross section for Ne~VII$\lambda$973.3 is marked with a cross symbol.
In the lower panel, the branching ratio of Raman scattering into the $2s$ state is shown by
a dotted line and the branching ratio into levels with $n=3$ is shown by a dot-dashed line.
 }
 \label{cross_section}
\end{figure}

In Fig.~\ref{cross_section} we show the scattering cross section and branching ratios 
near Ly$\gamma$ computed by \cite{jl04}. The upper panel shows the total scattering cross
section in units of the Thomson scattering cross section $\sigma_{Th}
=0.665\times10^{-24}{\rm\ cm^2}$. For $\lambda=973.302{\rm\ \AA}$
the total scattering cross section $\sigma_{tot}=394\sigma_{Th}=2.62\times 10^{-22}{\rm\
cm^2}$, which is
marked by a cross in the figure. In the lower panel, the dotted line shows the branching ratio
$\sigma_{2s}/\sigma_{tot}$
for Raman scattering into the optical channel, where $\sigma_{2s}$ is the cross
section for the final state of $2s$. Another branching ratio into $n=3$ levels
is shown by a dot-dashed line, which shows the ratio of $\sigma_{3s,3d}/\sigma_{tot}$
with $\sigma_{3s,3d}$ being the cross section for final states of $3s$ and $3d$.
We find the branching ratio of 0.168 for the scattering 
channel into the level $2s$ leading to formation of Raman scattered Ne~VII$\lambda$973.

For comparison, we note that
the total scattering cross section for He~II$\lambda$972
is $9.1\times 10^{-22}{\rm\ cm^{2}}$, which is larger than that of Ne~VII$\lambda$973
by a factor of 3.5. This is due to the fact that He~II$\lambda$972 is closer to
resonant Ly$\gamma$ than Ne~VII$\lambda$973 is. 
It is interesting to note
that the total scattering cross section is decreasing but the branching ratio
is slowly increasing as a function of wavelength locally around Ne~VII$\lambda$973.

\subsection{Monte Carlo Simulations}


Broad Balmer wings are
fairly common in symbiotic stars and V1016~Cyg is no exception.
\cite{lee00} proposed that these wings
are formed through Raman scattering of far UV continuum around Lyman series emission
lines. It has been also proposed that the hot tenuous fast stellar wind from the hot component
is responsible for the Balmer wings in these objects (e.g. \citealt{sk06}).
It is a difficult task to determine unequivocally the origin of the Balmer wings
for an individual source. However, in the case of V1016~Cyg, \cite{lee12}
proposed that the Raman scattering origin of the Balmer wings 
is quite plausible in his analysis of Raman scattered He~II$\lambda$4332. 

In this work we assume that the broad H$\beta$  wings 
are also formed through Raman scattering of far UV continuum around Ly$\gamma$.
%
To the far UV continuum we add the two emission components of He~II and Ne~VII
so that the input spectrum consists of the following 
three components, i.e., a locally flat continuum, and the two emission 
components of He~II$\lambda$972 and Ne~VII$\lambda$973. Both the emission components are prepared to exhibit a Gaussian profile given by Eq.~(\ref{gaussian}).
For the 2003 spectrum, both the He~II emission region and the Ne~VII$\lambda$973
region are assumed
to be at rest with respect to the H~I region.  

As is shown in the works of \cite{lee00} and \cite{jl04}, the Raman scattered 
radiation of far UV continuum around Ly$\beta$ is characterized by a plateau 
near the Balmer core
and an extended wing approximately proportional to $\Delta\lambda^{-2}$. 
The extent of the plateau is
determined by the scattering optical depth of a few, 
because the Raman conversion
efficiency becomes saturated above this scattering optical depth. 
Therefore the width of the plateau increases as $N_{HI}$ increases, 
which may be used to put a constraint on the value of $N_{HI}$.

\cite{jl04} proposed the neutral hydrogen column density $N_{HI}=1.21\times10^{21}{\rm\
cm^{-2}}$ in their analysis of Raman scattered He~II$\lambda$972 in V1016~Cyg. 
With a column density exceeding $N_{HI}=5\times 10^{21}{\rm\ cm^{-2}}$, Raman scattered He~II$\lambda$972
is formed in the plateau part of the H$\beta$ wing, which is inconsistent with our BOES data. 
On the other hand, if $N_{HI}$ is much
lower than $10^{21}{\rm\ cm^{-2}}$, then the scattering optical depths for both He~II$\lambda$972 and Ne~VII$\lambda$973 become so small, which would reduce Raman conversion efficiency significantly.

The same Monte Carlo code introduced by \cite{lee12} is used in the current
work, where the scattering region is assumed to be  a  uniform
sphere with the point-like far UV emission source located at the center.
The neutral hydrogen column density $N_{HI}$ is measured radially 
from the center to the edge of the sphere. The code is designed to take
full considerations of Rayleigh and Raman scattering channels faithful to the
atomic physics.  A given far UV photon generated in the emission region 
is treated to escape from the region immediately after its first Raman
scattering or reaching the spherical edge after a series of multiple Rayleigh 
scatterings.

\begin{figure}
\includegraphics[scale=0.65]{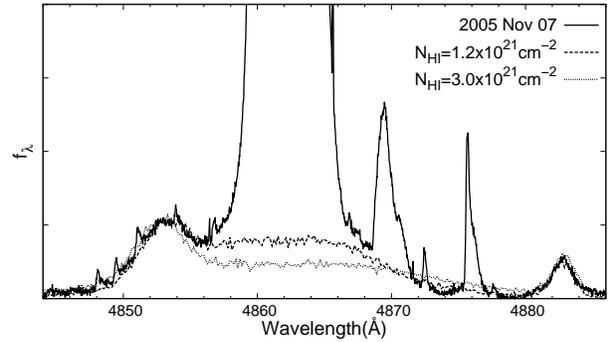}
 \caption{Monte Carlo simulations of Raman scattering of far UV continuum around Ly$\beta$ with the two emission components of He~II$\lambda$972 and Ne~VII$\lambda$973. The thick
 dashed line shows the result for $N_{HI}=1.2\times 10^{21}{\rm\ cm^{-2}}$ and the
 thin dashed line is for $N_{HI}=3.0\times 10^{21}{\rm\ cm^{-2}}$. In the latter case,
 the Raman scattered He~II feature is blended to the plateau region of Raman scattered
 far UV continuum, providing a poor fit to the observed data.
}
\label{col_dep}
\end{figure}

Fig.~\ref{col_dep} shows our Monte Carlo simulated profiles superposed to the 2005 BOES
data for two values of $N_{HI}$. In the case of $N_{HI}=3.0\times 10^{21}{\rm\ cm^{-2}}$,
the Raman scattered He~II feature sits near the edge of the plateau region, 
yielding a poor fit to the observed data.
In view of this result and for want of a better constraint, we adopt $N_{HI}=
1.21\times10^{21}{\rm\ cm^{-2}}$, as was proposed by \cite{jl04}, in order to
investigate the basic properties of Raman scattering of Ne~VII$\lambda$973.

\begin{figure}
\includegraphics[scale=0.65]{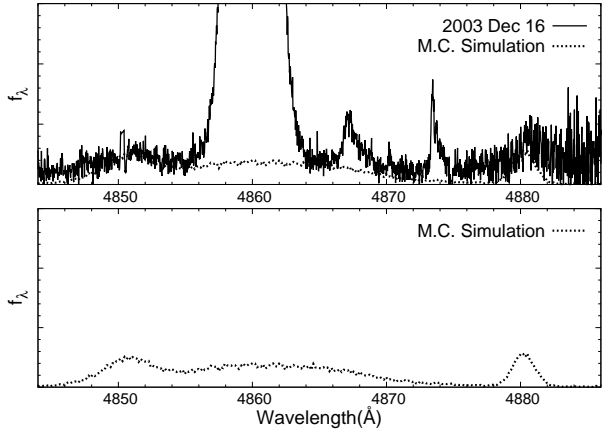}
 \caption{Monte Carlo simulated fit of Raman scattered He~II$\lambda$972 and Ne~VII$\lambda$973. 
 The adopted neutral hydrogen column density is
 $N_{HI}=1.2\times 10^{21}{\rm\ cm^{-2}}$. 
 In the upper panel, we show the 2003 data by a solid
 and the Monte Carlo result by a dotted line. 
In the lower panel, the BOES data is deleted 
in order to show clearly the Monte Carlo result. 
For the 2003 data, 
both the He~II and Ne~VII emission regions
 are assumed to coincide and be at rest with respect to the H~I region.
 }
 \label{03mc}
\end{figure}

In Fig.~\ref{03mc} we show our Monte Carlo result that appears 
to give the best fit to the 2003 BOES data. In the lower panel, 
only the Monte Carlo
simulated profile is shown by a dotted line. In the upper panel, 
the simulated profile
is superposed on the 2003 spectrum. We find the overall agreement 
of the Monte Carlo
profile and the BOES spectrum, although the observed Raman scattered Ne~VII$\lambda$973
is of poor quality. The best fitting Gaussian width for the incident 
He~II$\lambda$972
is $\Delta\lambda=0.10{\rm\ \AA}$. Defining the random velocity scale for He~II by 
\begin{equation}
v_{ran}\equiv {\Delta\lambda\over\lambda_0}c,
\end{equation}
we find $v_{ran}^{HeII}= 31{\rm\ km\ s^{-1}}$. Correspondingly for Ne~VII$\lambda$973
we have $\Delta\lambda=0.046{\rm\ \AA}$, which yields $v_{ran}^{NeVII}=14{\rm\ km\ s^{-1}}$,
smaller by a factor of $\sim 2$ than the He~II counterpart.
The equivalent widths of He~II$\lambda$972 and Ne~VII$\lambda$973 are
$EW_{He}=0.80{\rm\ \AA}$ and $EW_{Ne}=2.6{\rm\ \AA}$, respectively. However,
the equivalent widths are uncertain by a factor of 2,
considering the poor quality of the 2003 spectrum around the Raman features.

\begin{figure}
\includegraphics[scale=0.65]{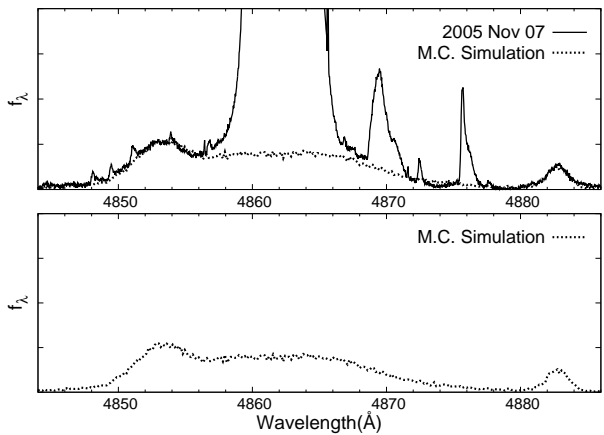}
 \caption{Monte Carlo simulated fit to the 2005 BOES data.  
The symbols are the same as Fig.\ref{03mc}. In this simulation, 
both the He~II and Ne~VII emission regions
 are assumed to coincide but recede from the H~I region with a velocity $v=+20{\rm\ km\
 s^{-1}}$.
 }
 \label{05nevii_sim}
\end{figure}

Similar Monte Carlo simulations with a modification of the two emission components in the input
spectrum are performed to fit the 2005 BOES data.
Fig.~\ref{05nevii_sim} shows our Monte Carlo result
for the best fit, from which we find that both the He~II and Ne~VII regions
recede from the H~I region with a velocity
$v=20 {\rm\ km\ s^{-1}}$.  The equivalent width of He~II$\lambda$972 for this case is
$0.5 {\rm\ \AA}$. Correspondingly for Ne~VII it is $1.4 {\rm\ \AA}$, which
means that Ne~VII$\lambda$973 is almost 3 times stronger than He~II$\lambda$972.
However, the poor quality of the 2003 spectrum prevents us from drawing any conclusion 
about the temporal change in the total line fluxes of He~II$\lambda$972 and
Ne~VII$\lambda$973 in the two year period. We summarize
the best fitting parameters in Table~3.

No significant change in the line widths of the Raman features are found in the two spectra. 
We simply assume that the random velocity scale $v_{ran}$ is contributed by a thermal 
component $v_{th}$ and 
a turbulent component $v_{turb}$ so that we may write
\begin{equation}
v_{ran}^2=v_{th}^2+v_{turb}^2.
\end{equation}
We further assume that the temperature and $v_{turb}$ are the same for both He~II and Ne~VII regions,
then the thermal velocity scale  $v_{th}^{He}$ of He~II will be larger than that $v_{th}^{Ne}$ 
of Ne~VII by a factor
$\sqrt{5}$. Using the fitted result we find
$v_{turb}=7{\rm\ km\ s^{-1}}$, $v_{th}^{Ne}= 12{\rm\ km\ s^{-1}}$, and $
v_{th}^{He}=27{\rm\ km\ s^{-1}}$, from which we find the temperature of the emission region 
$T=3.0\times 10^4{\rm\ K}$.

\begin{table}
\centering
\caption{Best Fitting Parameters from Monte Carlo Simulations}
\begin{tabular}{ccc} 
\hline
2003 Spectrum & He II 972 & Ne VII 973 \\ \hline
$\Delta\lambda$ & 0.10 \AA & 0.046 \AA \\ 
Equivalent Width  & 0.8 \AA & 2.6 \AA  \\ \hline
\hline
2005 Spectrum & He II 972 & Ne VII 973 \\ \hline
$\Delta\lambda$ & 0.10 \AA & 0.046 \AA \\ 
Equivalent Width  & 0.5 \AA & 1.4 \AA  \\ \hline

\end{tabular}
\end{table}

\section{Summary}

The operation of Raman scattering by atomic hydrogen is a very unique astrophysical
process that distinguishes symbiotic stars from other celestial bodies. Thus far
the Raman scattered line features found in symbiotic stars are limited to only He~II and O~VI.
In this work we propose that the broad feature around 4881 \AA\ in the
spectrum of V1016~Cyg is formed through Raman scattering of Ne~VII$\lambda$973.
We have provided the atomic physical quantities regarding Raman scattering of Ne~VII$\lambda$973. 
By performing Monte Carlo simulations, we obtain the random velocity scale of Ne~VII
and infer the equivalent width of the emission line Ne~VII$\lambda$973.

Assuming that both Ne~VII$\lambda$973 and He~II$\lambda$972 emission regions coincide,
we find that both He~II and Ne~VII emission components are
receding from the neutral region with the same speed $v\sim 20{\rm\ km \ s^{-1}}$ in the
2005 BOES data. 
However, the 2003 BOES data indicate that both Ne~VII and He~II 
emission regions are consistent with their being at rest with respect to the H~I region 
at that time.

\section*{Acknowledgements}

We are grateful to an anonymous referee for very useful comments 
that improved the presentation of the paper.
We also thank the staff of the Bohyunsan Optical Observatory for their help in
securing the high resolution spectra of V1016~Cyg. This research was 
supported by the Basic Science Research Program through the National 
Research Foundation (NRF)
funded by the Ministry of Education, Science and Technology (2011-0027069).

\bsp

\label{lastpage}


\begin{thebibliography}{99}

\bibitem[\protect\citeauthoryear{Angeloni et al.}{2010}]{an10} Angeloni, R., Contini, M.,
Ciroi, S., Rafanelli, P., MNRAS, 402, 2075
\bibitem[\protect\citeauthoryear{Belczynski et al.}{2004}]{bel00}  
Belczy\'nski, K., Mikołajewska, J., Munari, U., Ivison, R. J., Friedjung, M., 2000, A\&AS, 
146, 407
\bibitem[\protect\citeauthoryear{Bethe \& Salpeter}{1957}]{bs57}
Bethe, H. A. \& Salpeter, E. E., 1957, Quantum Mechanics of One- and
Two-Electron Atoms, New York, Academic Press
\bibitem[\protect\citeauthoryear{Birriel}{2004}]{bi04} Birriel J., 
2004, ApJ, 612, 1136
\bibitem[\protect\citeauthoryear{Edl\'en}{1983}]{ed83} Edl\'en, B., 1983, Physica Scripta,
28, 51
\bibitem[\protect\citeauthoryear{Feldman et al.}{1997}]{fe97}
Feldman U., Behring W. E., Curdt W., Sch\"ule U., Wilhelm K., Lemaire P. \& Moran K. T. M.,
1997, ApJ, 113, 195
\bibitem[\protect\citeauthoryear{Harries \& Howarth}{1997}]{hh97}
Harries, T. J.\&  Howarth, I. D., 1997, A\&AS,  121, 15
\bibitem[\protect\citeauthoryear{Herald et al.}{2005}]{he05}
Herald J. E., Bianchi L. \& Hillier D. J., 2005, ApJ, 627, 424
\bibitem[\protect\citeauthoryear{Iben \& Tutukov}{2004}]{ib96} Iben, I., Jr., Tutukov, A. V.,
ApJS, 105, 145
\bibitem[\protect\citeauthoryear{Jung \& Lee}{2004}]{jl04} Jung Y.-C. \& Lee H.-W., 
2004, MNRAS, 355, 221
\bibitem[\protect\citeauthoryear{Kaastra et al.}{1995}]{ka95} Kaastra, J. S., Roos, N.,
\& Mewe, R. 1995, A\&A, 300, 25,
\bibitem[\protect\citeauthoryear{Kenyon}{1986}]{ke86} Kenyon, S. J., 1986, 
The Symbiotic Stars, Cambridge University Press
\bibitem[\protect\citeauthoryear{Kramida et al.}{2006}]{kr06} Kramida, A. \&
Buchet-Poulizac, M.-C., 2006, European Physical Journal D, 38, 265
\bibitem[\protect\citeauthoryear{Kramida et al.}{2013}]{kr13} Kramida, A., Ralchenko, Yu., 
Reader, J., and NIST ASD Team, 2013, {\it NIST Atomic Spectra Database (ver. 5.1)} 
\bibitem[\protect\citeauthoryear{Lee}{2000}]{lee00} Lee H.-W., 2000, ApJ, 541, L25 
\bibitem[\protect\citeauthoryear{Lee}{2012}]{lee12} Lee H.-W., 2012, ApJ, 750, 127
\bibitem[\protect\citeauthoryear{Lee et al.}{2006}]{lj06} Lee H.-W., Jung Y.-C., Song I.-O. 
\& Ahn S., 2006, ApJ, 636, 1045
\bibitem[\protect\citeauthoryear{Lee \& Kang}{2007}]{lk07} Lee H.-W. \& Kang S., 
2007, ApJ, 669, 1156
\bibitem[\protect\citeauthoryear{Lee \& Park}{1999}]{lp99} Lee H.-W. \& Park M.-G., 
1999, ApJ, 515, L89
\bibitem[\protect\citeauthoryear{Luna et al.}{2013}]{lu13} Luna, G. J. M., Sokoloski, K.,
Mukai, K. \& Nelson, T., 2013, A\&A, 559, A6
\bibitem[\protect\citeauthoryear{Mastrodemos \& Morris}{1998}]{ma98} Mastrodemos, N.,
Morris, M., 1998, ApJ, 497, 303 
\bibitem[\protect\citeauthoryear{Mikolajewska}{2012}]{mi12} Mikolajewska, M., Baltic Astronomy,
21, 5
\bibitem[\protect\citeauthoryear{McCusky}{1965}]{mc65} McCusky, S., 1965, IAU Circ. 1916
\bibitem[\protect\citeauthoryear{M\"urset et al.}{1991}]{mu91} M\"urset, U. Nussbaumer, H., 
Schmid, H. M. \&
Vogel, M., 1991, A\&A, 248, 458
\bibitem[\protect\citeauthoryear{Nussbaumer}{2003}]{nu03} Nussbaumer, H., 2003, ASPC, 303, 557
\bibitem[\protect\citeauthoryear{Schmid}{1989}]{sc89} Schmid, H. M., 
1989, A\&A, 211, L31
\bibitem[\protect\citeauthoryear{Schmid}{1992}]{sch92} Schmid, H. M., 1992, A\&A, 254, 224  
\bibitem[\protect\citeauthoryear{Skopal}{2006}]{sk06} Skopal, A., 2006, A\&A, 457, 1003
\bibitem[\protect\citeauthoryear{Sokoloski et al.}{2001}]{sok01} Sokoloski, J. L.,
Bildsten, L., \& Ho, W. C. G., 2001, MNRAS, 326, 553
\bibitem[\protect\citeauthoryear{Torres et al.}{2012}]{to12} Torres A. F., Kraus M.,
Cidale L. S., Barb\'a R., Borges Fernandes M. \& Brandi E., 2012, MNRAS, 427, L80
\bibitem[\protect\citeauthoryear{van Groningen}{1993}]{vg93} van Groningen, E., 
1993, MNRAS, 264, 975
\bibitem[\protect\citeauthoryear{Warner}{1995}]{wa95} Warner, B., 1995, Cataclysmic Variable
Stars, Cambridge University Press,
\bibitem[\protect\citeauthoryear{Werner et al.}{2004}]{we04} Werner, K., Rauch, T.,
Reiff, E., Kruk, J. W. \& Napiwotzki, R., 2004, A\&A, 427, 685
\bibitem[\protect\citeauthoryear{Young et al.}{2005}]{yo05} Young, P. R.,
Dupree A. K., Espey B. R., Kenyon, S. J., Ake, T. B., 
2005, AJ, 618, 891

\end{thebibliography}
\end{document}